# Low beta spoke cavity multipacting analysis

XU Bo(徐波)[1] LI Han(李菡)[1,2] ZHANG Juan(张娟)[1,2] SHA Peng(沙鹏)[1] WANG Qunyao(王群要)[1] LIN Haiying(林海英)[1]

HUANG Hong(黄泓)[1] DAI Jianping(戴建枰)[1] SUN Yi(孙毅)[1] WANG Guangwei(王光伟)[1] PAN Weimin(潘卫民)[1]

1 (Institute of High Energy Physics, CAS, Beijing 100049, China)

2 (University of Chinese Academy of Sciences, Beijing 100049, China)

**Abstract:** The simulation and analysis for electron multipacting phenomenon in low β spoke superconducting cavity in ADS proton accelerator are proposed. Using both CST and Track3P codes, the electron multipacting calculation for β = 0.12 spoke superconducting cavity is implemented. The methods of multipacting calculation on both codes are studied and described. With the comparison between the calculation results and the cavity vertical test result, the accuracy and reliability of different code on calculating multipacting are analyzed. Multipacting calculation can help to understand the result of vertical test and also can help to do the optimization in cavity design.

**Key words:** ADS, superconducting cavity, secondary emission coefficient, multipacting

**PACS:** 29.20.Ej

## 1. Introduction

Spoke cavity is widely adopted in proton accelerator for its advantages for low and medium β particles acceleration. A type of 325MHz β=0.12 spoke cavity (spoke012 mentioned below) was adopted in Injector I of China Accelerator Driven Sub-critical System (C-ADS) project as shown in Fig.1 [1-2]. Multipacting is an issue of concern in the design of superconducting resonator. Some multipacting bands which are called soft barriers can be conditioned and eliminated with RF, while hard multipacting barriers may prevent the resonators from reaching the design voltage [3] and affect the performance of cavity. For instance, the SSR1 spoke cavity in Fermilab encountered strong multipacting during its test [4] and multipacting has become a main obstacle to improve the gradient of the cavity. It is necessary to do some research on multipacting especially on spoke cavity design.

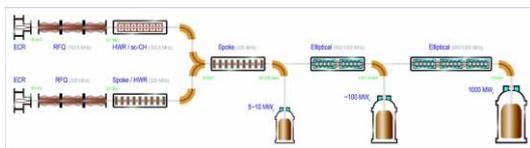

Fig.1. The Linac layout of ADS project

With the development of computer technology, the study of the multipacting can be executed in detail by using large scale computer. By means of the calculation and simulation of the multipacting, on the one hand superconducting cavity can be optimized by changing shape, to reduce the occurrence possibility of the multipacting, on the other hand multipacting codes can benefit from the cavity test in further development and can be improved on the accuracy and reliability.

## 2. The codes and algorithm of multipacting calculation

Multipacting simulation provides an effective method to study the Multipacting effects. Normally a reliable multipacting calculation require the ability of code as following [5]:

1) Accurate electromagnetic field;
2) Suitable particle emission model;
3) Proper SEC (Secondary Emission Coefficient) curve for surface material;
4) Comprehensive postprocessing of particle data.

At present, two codes for multipacting calculation are very popular, one is Track3P module from ACE3P （ Advanced

Computational Electromagnetic 3D Parallel code ）which is developed by SLAC [6], and the other one is PS（Particle Studio）module from CST [7], which also has the ability for the calculation of secondary electron emission.

The multipacting happens when the conditions as follow are satisfied in cavity:

1) Initial electrons are released from the surface of cavity.
2) Secondary emitted electrons are in resonance with the RF fields.
3) Impact energies of the secondary electrons fall within the range of SEC greater than unity. SEC means the number of secondary electrons emitted per incident particle.
4) The number of resonant electrons multiplies exponentially, leading to a phenomenon of electron avalanche.

In CST code, the secondary emission coefficient as shown in Fig.2 can be set combined with material. The SEC can be considered in the procedure from the initial electrons hit the surface to secondary electrons emit from the surface. Thus we can obtain a curve of electron number increasing or decreasing with time. By increasing the amplitude and changing the phase of EM field step by step, a curve describing the growth rate of electron numbers vs. accelerating gradient (Eacc) can be obtained.

In Track3P code, according to the final data of the resonant particle, we can plot out a chart of resonant particle energy and accelerating gradient (Eacc). We use different secondary emission coefficient to evaluate the multipacting probability. For instance we pick the SEC of Niobium 300 degree bake out, as shown in Fig.2. According to the value of SEC which is greater than unity, we can obtain the multipacting range of kinetic energy which is from about 70eV to 1500eV. Using the above energy range to cut the resonant energy of the chart, the according data of different Eacc is concerned. If the data in the according energy range is dense, the probability of the multipacting occurrence may be bigger otherwise the probability may be smaller. In general we can evaluate the multipacting level by analyze the density of data from specified resonant energy and specified accelerating gradient.

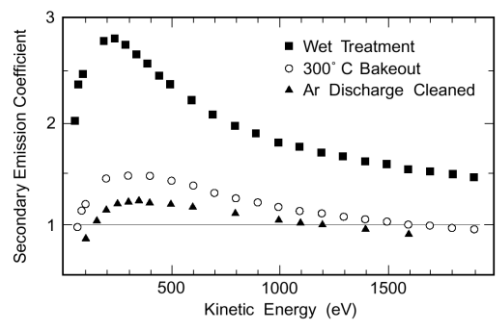

Fig.2. The secondary Emission Coefficient of Niobium

## 3. The electromagnetic field distribution of spoke012 cavity

The driven electromagnetic field calculation is important to the following multipacting calculation. At present, most of electromagnetic field calculation codes can implement 3D model structure calculation. We can set up the cavity model in certain code preferred and import the model file to the electromagnet calculation code. In CST code, we use Microwave Studio (MWS) to set up a model and use Eigenmode solver to implement the electromagnet calculation. In ACE3P code, we use Omega3P module to implement the electromagnetic field calculation. The electromagnetic field distribution result of the spoke012 cavity from Omega3P module is shown in Fig.3. For the mesh cell setting, one difference is Hexahedral cells are used in CST while Tetrahedral cells are used in Omage3P. Obviously the more mesh cells used, the more accurate the EM field result is. But at the same accuracy class, the number of high order tetrahedral cell is much less than the number of hexahedral cells.

It is better to use curved tetrahedral cell in EM calculation.

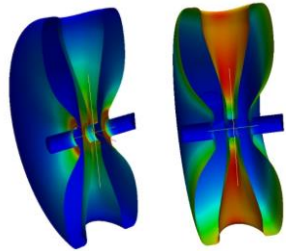

Fig.3. The electromagnetic field distribution of spoke012
(Left: electric field, right: magnetic field)

## 4. The spoke012 multipacting calculation result from CST

Multipacting calculation can be executed after electromagnetic field calculation. Multipacting calculation is carried out using CST Particle Studio. A metal shell has to be made to cover the vacuum solid model. Different SEC curve can be chosen for the metal shell. For niobium, the curve of 300 centigrade bake out was used in our calculation. On the inner surface of metal shell, some suspicious faces can be chosen as primary electrons source. The surface will be divided by certain number grids and the primary electrons will emit from the center of grid. The EM field can be imported then the electrons which emit from the inner surface are driven by the EM field, thus the motion of the electrons can be calculated. Fig.4 shows a result of particle numbers increasing with time when multipacting happened. Different curves can be obtained as we change the amplitude and phase of EM field step by step. Thus the slope value of every curve can be obtained. We can convert the amplitude value to be as accelerating gradient and we chose the biggest slope value from different phase at the same accelerating gradient to be as a mark point of growth rate value. A curve of growth rate vs. Eacc can be obtained as shown in Fig.5 by the principle above. From the curve we can see that three peak of multipacting appeared corresponding to the accelerating gradient 2.5MV/m, 4.5MV/m and 8.7MV/m. When Eacc is higher than 9.5MV/m, there will be no multipacting happened.

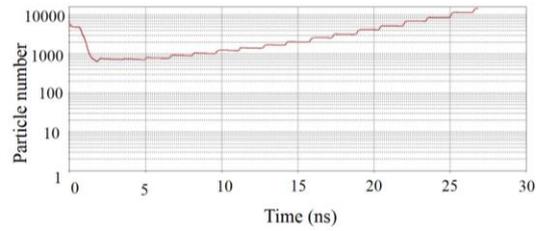

Fig.4. Particle number vs. time

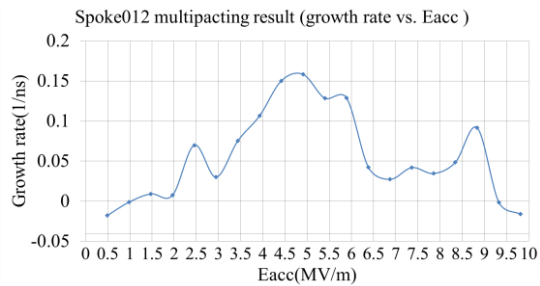

Fig.5. The spoke012 multipacting result curve of growth rate vs. Eacc from CST

## 5. The spoke012 multipacting calculation result from Track3P

The ACE3P code suite is based on the Finite Element Method (FEM) using tetrahedral curved elements for accelerator computation developed by SLAC. Track3P module from ACE3P has been used to do the multipacting calculation in many cavities [8]. ACE3P runs on a super computer so it can do a large scale parallel computation. We use Omega3P to do the EM calculation as mentioned above then use track3P to simulate the multipacting in spoke012 cavity. Track3P is some different from CST PS. Firstly, the mesh cells for EM field calculation are different. The mesh cells for EM calculation in CST PS can only be set as hexahedral cells while in Omega3P from ACE3P the cells can be set as 2 order tetrahedral cells. Normally high order tetrahedral cells can have a better approximation to curved cavity surface. Secondly, Track3P calculates the particle trajectories and resonant energy left after specified RF periods no matter what the material is for the cavity shell. After the resonant energies on every accelerating gradient are obtained, we can use the SEC curve of different material to delimit the energy range of multipacting. The concerned resonant energy range is as the same incident kinetic energy range corresponding to SEC value greater than unity. For Niobium of 300 centigrade bake out, the energy range is about from 70eV to 1600eV. We plotted out the resonant energy of every resonant particle left after 50 RF periods on every scanned accelerating gradient as shown in Fig.6. The positions where the resonant particles located were also plotted out in Fig.6. From the result we can conclude that multipacting may happen at Eacc around 3MV/m, 5MV/m and 8MV/m. The locations where the multipacting happened are transferred from spoke top to coupler port with the gradient increasing. When the accelerating gradient is greater than 9MV/m, there will be no multipacting happened.

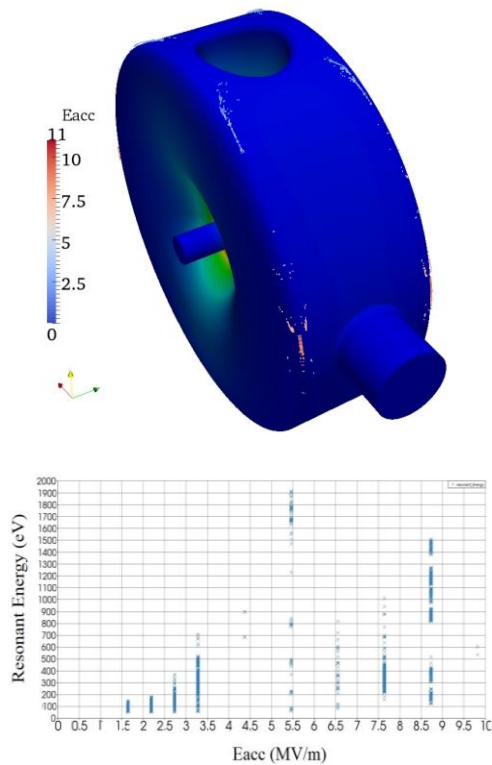

Fig.6. Spoke012 multipacting location (top) and the result of multipacting resonant energy vs. Eacc (bottom) fromTrack3P

## 6. The spoke012 vertical test result and multipacting result analysis

By means of doing the vertical test, the performance of spoke012 cavity was verified. The photo of spoke012 cavity (left) and the cavity fixed on the support stick before Vertical Test (right) are shown in Fig.7. The vertical test data $Q_0$ and Eacc of spoke012 is shown in Fig.8.

In the vertical test of spoke012, some multipacting happened at certain $E_{acc}$ which were also released as data shown in Fig.8. As mentioned above, multipacting normally can be divided into soft barriers and hard barriers. Soft barriers can be eliminated by RF power in the process of conditioning while hard barriers are not easy to be removed and can be a trouble for cavity. We met soft barriers multipacting at about 2.5MV/m and 5MV/m which caused $Q_0$ dropped at the first time of $Q_0$ measurement. With input power increasing and conditioning, $Q_0$ increased at these gradient points and $Q_0$ curve became smooth. The soft barriers multipacting was easy to be removed at the accelerating gradient mentioned above.

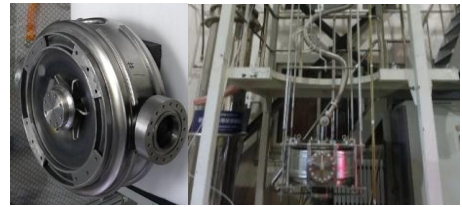

Fig.7. Spoke012 cavity (left) and spoke012 cavity installed in vertical test stand (right)

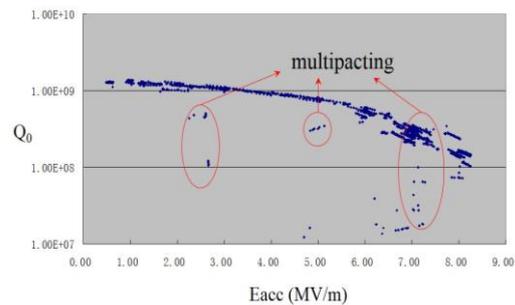

Fig.8. The vertical test data $Q_0$ vs. $E_{acc}$ of spoke 012

From the data in Fig.8, we can find that $Q_0$ dropped a lot of times when Eacc is around and above 7MV/m. At the same time, the $Q_0$ curve became not smooth as before and the value measured was not stable. It had no any improvement on $Q_0$ value even after strong RF conditioning for long time. Judged from the experience and phenomenon, it might be a hard multipacting barrier. The hard barrier may lead to the decreasing of $Q_0$ value and confusion of $Q_0$ measurement just like SSR1 cavity in Fermilab. The hard barrier multipacting may be one reason for the performance dropping of spoke012 cavity.

## 7. The analysis for the result from simulation and test

CST PS and Track3P from ACE3P were used to analyze the multipacting in spoke012 cavity. Compare the simulation results from two codes, we can obtain something in common and something in different. The point in common is that there will be no multipacting when $E_{acc}$ is greater than 10MV/m. This means it should be no multipacting in high accelerating gradient according to this shape of spoke012. In low accelerating gradient, multipacting are calculated out from both results of different codes, but the $E_{acc}$ values for multipacting point are not same exactly. From the result of CST PS, the multipacting peak point is about 5MV/m, while the result of Track3P shows the multipacting peak point is about 8-9MV/m. From the vertical test result presented in Fig.8, we found one multipacting peak point around 7-8MV/m while multipacting on 3MV/m and 5MV/m are soft barriers. What is the reason for the difference between the results of two codes? From our opinion, one reason may be the calculation capacity of different codes. In Track3P computation, hundreds of CPU and large memories are used to implement the task while in CST the computing resource is less. The other reason may be the different of mesh density. Due to the computation capacity difference, the mesh cells in Track3P are much more than the cells in PS.

Due to the reason about the difference between the simulation result and test result, it is still a complex and uncertain problem. Multipacting is a complex phenomenon in cavity anyway, it depends on so many factors such as the cleanness of the cavity surface and the EM field distribution on cavity inner surface and so on. It is hard to obtain a very accurate result just from the simulation and even the multipacting tendency is difficult to be obtained in 3D model structure before. Now the result of multipacting simulation can be calculated and it can be verified by test result on a certain extent mostly. On the other hand, the simulation result can also be an important reference in process of cavity design and be a good guide in process of test. We can evaluate the multipacting tendency to optimize cavity shape. In process of the test, some measures can be done to avoid or overcome multipacting refer to the simulation result. For example, we can arrange power conditioning according to the simulation result in vertical test, we can avoid the multipacting range to do some measurement. All these cannot be implemented in the past. Multipacting simulation makes the test not 'blind' any more.

## 8. Summary

We have done the multipacting simulation for spoke012 using CST PS and Track3P. Vertical test on spoke012 has also been implemented. The test result verifies the simulation result and also benefits from the simulation result. In the future how to make the simulation more accurate is still a problem to be studied and how to identify the soft barrier and hard barrier from the simulation is a further problem to be solved.

*We would like to thank the Advanced Computations Department of SLAC for providing ACE3P code suite and computing resources.*